\documentclass{article}

\usepackage{arxiv}

\usepackage[utf8]{inputenc} % allow utf-8 input
\usepackage[T1]{fontenc}    % use 8-bit T1 fonts
\usepackage{hyperref}       % hyperlinks
\usepackage{url}            % simple URL typesetting
\usepackage{booktabs}       % professional-quality tables
\usepackage{amsfonts}       % blackboard math symbols
\usepackage{nicefrac}       % compact symbols for 1/2, etc.
\usepackage{microtype}      % microtypography
\usepackage{lipsum}		% Can be removed after putting your text content
\usepackage{graphicx}
\usepackage{natbib}
\usepackage{doi}
\usepackage{multirow}  % used for merge of comparison tables
\usepackage{graphicx}  % userd for vertical columns
\usepackage{tabularx}
\usepackage{amsmath}
\usepackage{tabularx}
\usepackage{threeparttable}
\usepackage{comment}
\title{Vocal Prognostic Digital Biomarkers in Monitoring Chronic Heart Failure: A Longitudinal Observational Study} 

\author{
\begin{tabular}{c}
\bfseries
Fan Wu$^{a,b}$\thanks{Corresponding author: fanwu@ethz.ch, fbarata@ethz.ch},
Matthias P.~Nägele$^{c}$, Daryush D.~Mehta$^{d,e,f,g}$ \\
Elgar Fleisch$^{a,b,h}$, Frank Ruschitzka$^{c}$, Andreas J.~Flammer$^{c}$, 
Filipe Barata$^{a}$\footnotemark[1] \\[6pt]
\normalfont\small
$^{a}$Centre for Digital Health Interventions, ETH Zurich, Zurich, Switzerland \\
\normalfont\small
$^{b}$Agentic System Lab, ETH Zurich, Zurich, Switzerland \\
\normalfont\small
$^{c}$Cardiology, University Heart Center Zurich, University Hospital Zurich, Zurich, Switzerland \\
\normalfont\small
$^{d}$Center for Laryngeal Surgery and Voice Rehabilitation, Massachusetts General Hospital, Boston, MA \\
\normalfont\small
$^{e}$Department of Surgery, Harvard Medical School, Boston, MA \\
\normalfont\small
$^{f}$Speech and Hearing Bioscience and Technology, Division of Medical Sciences, Harvard Medical School, Boston, MA \\
\normalfont\small
$^{g}$MGH Institute of Health Professions, Boston, MA \\
\normalfont\small
$^{h}$Centre for Digital Health Interventions, University of St.\ Gallen, St.\ Gallen, Switzerland
\end{tabular}
}

\renewcommand{\shorttitle}{Vocal Prognostic Biomarkers for Heart Failure}

\begin{document}
\maketitle

\begin{abstract}

\textbf{Objective: }
This study aimed to evaluate which voice features can predict health deterioration in patients with chronic HF.

\textbf{Background: }
Heart failure (HF) is a chronic condition with progressive deterioration and acute decompensations, often requiring hospitalization and imposing substantial healthcare and economic burdens. Current standard-of-care (SoC) home monitoring, such as weight tracking, lacks predictive accuracy and requires high patient engagement. Voice is a promising non-invasive biomarker, though prior studies have mainly focused on acute HF stages.

\textbf{Methods: }
In a 2-month longitudinal study, 32 patients with HF collected daily voice recordings and SoC measures of weight and blood pressure at home, with biweekly questionnaires for health status. Acoustic analysis generated detailed vowel and speech features. Time-series features were extracted from aggregated lookback windows (e.g., 7 days) to predict next-day health status. Explainable machine learning with nested cross-validation identified top vocal biomarkers, and a case study illustrated model application.

\textbf{Results: }
A total of 21,863 recordings were analyzed. Acoustic vowel features showed strong correlations with health status. Time-series voice features within the lookback window outperformed corresponding standard care measures, achieving peak sensitivity and specificity of 0.826 and 0.782 versus 0.783 and 0.567 for SoC metrics. Key prognostic voice features identifying deterioration included delayed energy shift, low energy variability, and higher shimmer variability in vowels, along with reduced speaking and articulation rate, lower phonation ratio, decreased voice quality, and increased formant variability in speech.

\textbf{Conclusion: }
Voice-based monitoring offers a non-invasive approach to detect early health changes in chronic HF, supporting proactive and personalized care.

\end{abstract}

% keywords can be removed
\keywords{
Heart failure \and Voice biomarker \and Prognostic biomarker \and Home monitoring \and Non-invasive \and Explainable AI
}

%%==================================%%
%% INTRODUCTION
%%==================================%%
\section{Introduction}
\label{articlec:sec:introduction}

Heart failure (HF) is a global health issue affecting an estimated 64.3 million people, characterized by the heart’s reduced ability to pump or fill with blood, leading to inadequate cardiac output~\cite{savarese2022global}. Chronic heart failure is a long-term condition that can progressively worsen and lead to acute decompensated HF, a sudden worsening of HF symptoms, often requiring hospitalization. Most acute HF hospitalizations are due to deterioration of existing chronic HF, with only 15–20\% representing new HF diagnoses~\cite{joseph2009acute}. These hospitalizations are a major driver of HF-related healthcare costs, accounting for approximately 75\% of total expenses, with each admission averaging 10,000€~\cite{bundkirchen2004epidemiology}. About 25\% of patients are readmitted within 30 days of discharge, and up to 75\% of these early readmissions may be preventable~\cite{joynt2011has,desai2012rehospitalization}. However, predicting readmissions is difficult due to the complex interplay of clinical, physiological, and social factors~\cite{desai2012rehospitalization}. The transition from hospital to home is a critical period when patients face an increased risk of early readmission, but it also offers a key opportunity for improving outcomes. Regular home monitoring of HF helps detect early signs of deterioration and timely intervention to reduce hospitalizations.

Current standard-of-care (SoC) home monitoring practices include daily weight and blood pressure monitoring. Weight gain from fluid retention can signal worsening HF~\cite{lyngaa2012weight}, and guidelines recommend monitoring thresholds (e.g., an increase of 1.5–2~kg/day) to prompt clinical action~\cite{senarath2021influential, mullens2024dietary}. However, weight gain often occurs too late and has low sensitivity (20–30\%) for predicting hospitalization~\cite{gyllensten2016early}. Blood pressure is another prognostic indicator, with chronic hypertension (high blood pressure) increasing the risk of HF development and progression~\cite{levy1996progression, ponikowski2019heart}; yet, in established HF, lower blood pressure can paradoxically indicate worse survival, and optimal targets and management remain unclear~\cite{maeda2023blood}. Moreover, SoC often demands active patient involvement, leading to poor and selective adherence~\cite{seid2019adherence}. There is a clear need for a novel solution that is passive, low-cost, convenient, accurate, and well-suited for remote home monitoring.

Voice, produced as air passes through partially closed vocal folds, is an emerging non-invasive biomarker linked to neurological, laryngeal, and other disorders, and has recently gained interest in cardiology~\cite{bauser2024voice}. Vocal biomarkers have been linked to hospitalization and mortality in patients with HF~\cite{maor2020vocal}. Studies have explored various voice features for HF assessment: Murton et al. associated weight loss with voice changes during hospitalization for acute decompensated HF~\cite{murton2017acoustic}; Amir et al. observed speech changes in 88\% of patients with acute HF between admission and discharge~\cite{amir2022remote}; Reddy et al. found mel-frequency cepstral coefficients (MFCC) features accurately differentiated healthy individuals from patients with acute HF~\cite{reddy2021automatic}; Firmino et al. used spectral and prosodic features to distinguish HF from healthy controls~\cite{firmino2023heart}; Schöbi analyzed speech and pause patterns in patients with acute HF and chronic HF~\cite{schobi2022evaluation}; and Zawa et al. linked maximum phonation time with exercise capacity and disease severity in chronic HF~\cite{izawa2015longitudinal}.

Earlier studies largely focused on distinguishing healthy individuals from patients with HF or differentiating between the acute states of patients with HF (e.g., at admission vs. discharge). There is a need to focus on patients with chronic HF to identify decompensation early—before clinical deterioration occurs, enabling timely interventions to reduce hospitalizations. Moreover, most existing studies are cross-sectional, providing only a snapshot in time. Longitudinal research is needed to capture dynamic changes and better predict worsening conditions. Previous studies mainly used professional voice recording equipment, but for voice to be a practical home monitoring tool, mobile device voice recording must be explored. Smartphone-based voice assessment and vocal biomarkers offer a noninvasive, low-cost, and easily accessible alternative, potentially enhancing remote patient management through mobile health solutions~\cite{fagherazzi2021voice}.

\begin{figure}[htbp]
    \includegraphics[trim={0 6cm 3cm 0},clip,width=\linewidth]{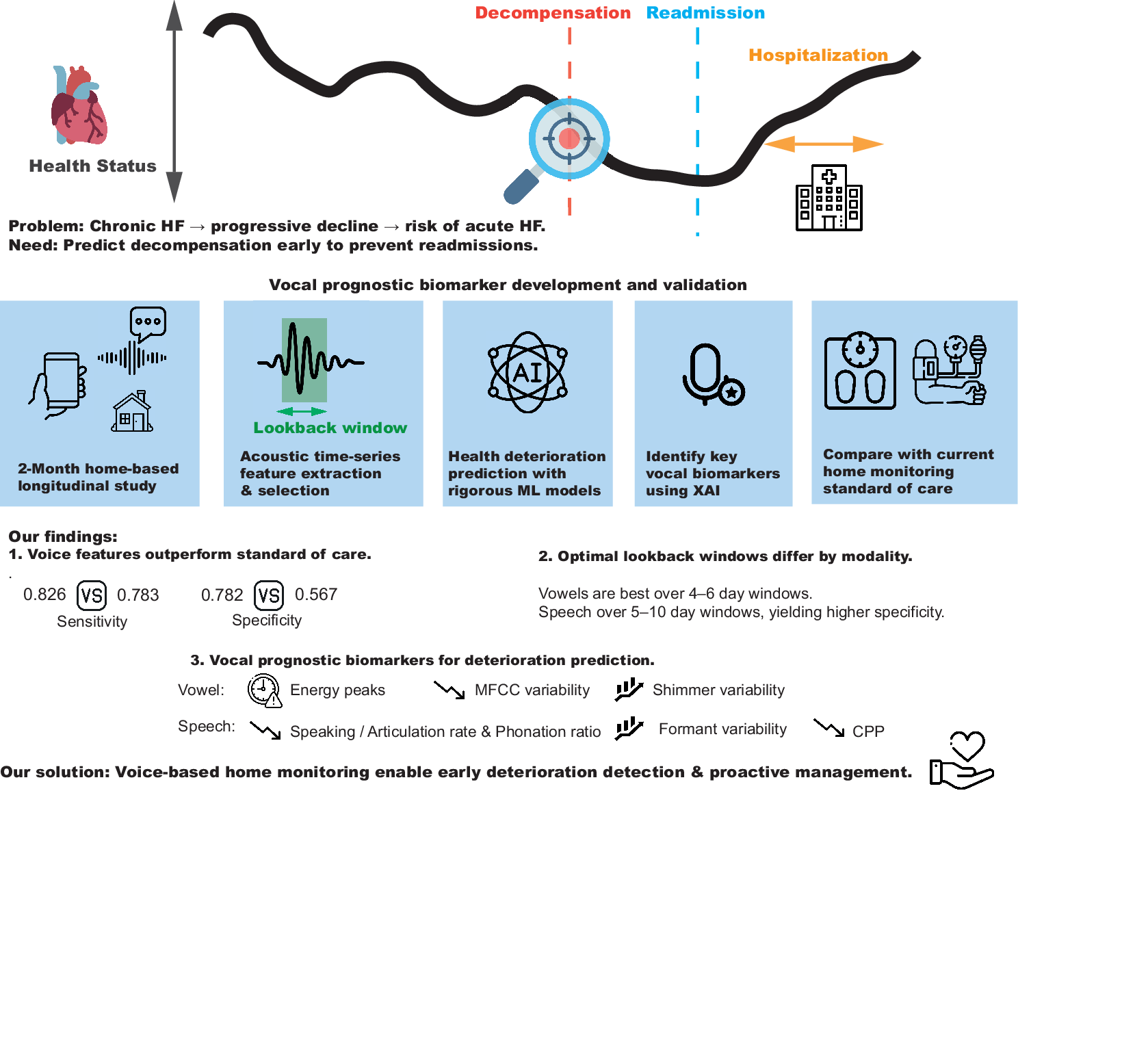}
    \centering
    \caption{\textbf{Voice-based monitoring system for chronic HF}. We conducted a home-based longitudinal study in which patients recorded daily speech and vowels. To develop and validate vocal biomarkers, we designed a pipeline including acoustic feature extraction, time-series analysis, and explainable machine-learning prediction of health deterioration, resulting in the identification of novel predictive biomarkers. Voice features outperformed SoC measures, highlighting their potential for earlier detection and proactive management to reduce hospitalizations.}
    \label{fig:fig0}
\end{figure}

To the best of our knowledge, this is the first study to examine the predictive power of vocal features for chronic HF and the first to conduct a long-term, home-based longitudinal study using mobile devices. Figure~\ref{fig:fig0} presents the central illustration of our voice-based monitoring system. We investigated how fluctuations in voice features over time can predict health status in patients with chronic HF in real-world settings. We compared voice with SoC measures, i.e., weight and blood pressure, to evaluate whether voice adds predictive value. Using data-driven methods, we included a wide range of acoustic features and identified those most predictive of health status. We also determined the optimal number of days of voice recordings required for accurate prediction. This work demonstrates the feasibility and effectiveness of voice as a prognostic biomarker and a potential self-management tool for chronic HF. To advance research, we release our rigorous, scalable, and explainable machine learning pipeline for acoustic feature engineering, model development, and interpretation as open source at \href{https://github.com/ADAMMA-CDHI-ETH-Zurich/CHFVoice}{GitHub}.

%%==================================%%
%% METHODS
%%==================================%%
\section{Methods}
\label{articlec:sec:methods}

\subsection{Study Design and Data Collection}

We conducted a two-month prospective longitudinal study in patients with chronic HF with fluctuating health status, approved by the Cantonal Ethics Committee Zurich (No. 2022-00912). Study design, eligibility criteria, and data collection procedures are detailed in our previous publication~\cite{wu2023trends}. To increase recruitment, the N-terminal pro B-type natriuretic peptide screening threshold was lowered to $\geq$125 pg/mL, with ethics approval. Patients used mobile devices for at-home data collection, recording 12 daily audio files, three vowels (/a/, /o/, /i/) and three speech tasks: the Voice Conversational Agent commands, the Northwind passage, and a 20-second narrative. SoC measures were obtained by patients who self-reported their daily weight and blood pressure. Daily symptoms were tracked using the Heart Failure Symptom Tracker (HFaST)~\cite{lewis2019development}, and biweekly health status was assessed with the Kansas City Cardiomyopathy Questionnaire (KCCQ), a validated 23-item tool covering domains such as physical limitation and symptom burden~\cite{green2000development}. The KCCQ overall summary score served as the ground-truth measure of health status. 

\subsection{Acoustic Feature Extraction}

All audio recordings were preprocessed to remove silence, low-frequency noise, and outliers. Silence was detected using an empirically determined root-mean-square threshold ($-40\,\mathrm{dB}$ re max), and recordings were high-pass filtered at $70\,\mathrm{Hz}$ to reduce microphone artifacts. Vowel recordings were further low-pass filtered at $3\,\mathrm{kHz}$, and voiced frames were identified using pitch-based voice activity detection. Vowels were min--max normalized to reduce variability. All recordings were manually inspected to remove extraneous sounds such as background noise (e.g., television) or mispronunciations.

We extracted acoustic features from vowel and speech recordings using three Python libraries: OpenSMILE~\cite{eyben2010opensmile}, SenseLab~\cite{senselab2025} and DisVoice~\cite{belalcazar2016glottal}. Vowel features included vocal-source measures (e.g., jitter, shimmer, harmonics-to-noise ratio), vocal-tract measures (e.g., formant frequencies, MFCCs, spectral features), and prosodic measures (e.g., pitch, energy). Low-level descriptors (LLDs) were extracted frame by frame from voiced segments identified by the voice activity detection method, with only recordings longer than 1 s retained. Statistical functionals (e.g., mean) were then computed for each vowel LLD and averaged across three daily repetitions, resulting in 5,254 functional features per vowel per day. For speech recordings, 25 features were extracted across timing, articulation, phonation, and respiration. An overview of the vowel LLDs and speech features is provided in Table~\ref{tab:acoustic_features}.

\begin{table}[htbp]
\centering
\begin{threeparttable}
\centering
\caption{Baseline Demographics and Clinical Characteristics (N = 32)}
\label{tab:acoustic_features}
\small

\begin{tabularx}{\textwidth}{p{5cm} X}

\hline
% Use {2}{l} to span both the 'l' and 'X' columns
\multicolumn{2}{l}{\textbf{Vowel LLD features}} \\ 
\hline
\textit{Feature Type} & \textit{Features} \\

Vocal-Source-Related LLDs (Phonation / Voice Quality)
& Log HNR; jitter (local); shimmer (local); probability of voicing; glottal features (e.g., open quotient, amplitude quotient) \\

Vocal-Tract-Related LLDs (Articulation / Spectral)
& MFCC 1--14; CPP; formants (F1, F2); spectrum bands 1--26 (0--8\,kHz); spectral energy 250--650\,Hz, 1--4\,kHz; spectral roll-off points 0.25, 0.50, 0.75, 0.90; spectral flux, centroid, entropy, slope, variance, skewness, kurtosis; psychoacoustic sharpness, harmonicity \\

Prosodic LLDs
& Sum of auditory spectrum (loudness); F0 (pitch); RMS energy; zero-crossing rate \\

\hline
\multicolumn{2}{l}{\textbf{Speech features}} \\
\hline
\textit{Feature Type} & \textit{Features} \\

Timing
& Speaking rate; articulation rate; pause rate; mean pause duration; phonation ratio \\

Articulation
& Formants and bandwidth: mean and standard deviation of F1 and F2; spectral moments: mean spectral gravity, standard deviation, skewness, kurtosis \\

Phonation (Voice Quality)
& Mean HNR; spectral slope; spectral tilt; CPP \\

Respiration (Prosody / Intensity)
& Mean pitch; standard deviation of pitch (semitones); mean intensity; intensity range \\

\hline
\end{tabularx}

\begin{tablenotes}
\footnotesize
\item HNR: harmonics-to-noise ratio; CPP: cepstral peak prominence; F1: first formant; F2: second formant.
\end{tablenotes}
\end{threeparttable}
\end{table}

\subsection{Repeated Measures Correlation}

The primary outcome, KCCQ, was collected biweekly, while voice, weight, blood pressure, and HFaST data were recorded daily. To align modalities, daily features were aggregated over K-day lookback window (K=2-14) preceding each KCCQ, using six time-series descriptors (mean, standard deviation, slope, rolling variability, and dominant spectral frequencies) to capture temporal dynamics for predictive modeling.

Given the longitudinal design, we calculated repeated-measures correlations~\cite{bakdash2017repeated} between KCCQ scores and each aggregated descriptor to capture within-subject associations. Based on these correlations, the top 20–30 vowel and speech features per time-series descriptor were retained, reducing dimensionality while preserving features most strongly associated with health status.

\subsection{Classification}

To evaluate which time-series features distinguished health status, we performed binary classification between the healthiest (KCCQ $> 87.5$) and most compromised groups (KCCQ $\leq 65.6$)~\cite{kosiborod2020effects}. Six feature sets were tested: (1) SoC, (2) HFaST, (3) Vowel, (4) Speech, (5) Voice (vowel + speech), and (6) all features. For each feature set, age, sex, and native language were included as covariates.

Random Forest and Extreme Gradient Boosting models were trained with recursive feature elimination for feature selection~\cite{darst2018using} and evaluated using subject-wise nested cross-validation~\cite{ghasemzadeh2024toward}. Performance was reported as mean ± SD across outer folds using Sensitivity, Specificity, F1 score, Matthews Correlation Coefficient (MCC), Area Under Receiver Operating Characteristic Curve (AUC), and Area Under Precision–Recall Curve (AUPRC). Feature importance was assessed with SHapley Additive exPlanations (SHAP) values~\cite{lundberg2017unified} to identify potential prognostic biomarkers. The analysis was further extended to a one-vs-one multi-class framework including moderate health status (see Supplementary Appendix).

\subsection{Statistical Analysis and Case Study}

We compared the distribution of the top voice features between the healthiest and most compromised health groups. Student's $t$-test was used for normally distributed features, and the Mann--Whitney $U$ test was applied for non-normal features. A two-sided significance level of $\alpha = 0.05$ was used, and effect sizes (Cohen's $d$ or rank-biserial correlation) were reported alongside $p$-values. Statistical analyses were performed in Python (version~3.10.15) using the SciPy library (version~1.15.2).

We conducted a case study on a patient who developed acute HF, whose data were fully withheld during model training and evaluation. Using a representative aggregation window, we performed binary classification between worst and best health states. The patient’s data were evaluated with the best-performing models, and average predicted probabilities of worst health were reported for each feature set.

\section{Results}

\subsection{Data Completeness and Retention Quality}

Patient recruitment ran from September 2022 to September 2024, enrolling 36 participants. Two missed technical visits and two withdrew, leaving 32 patients who completed the study. The demographics are summarized in Table~\ref{tab:clinical}. One patient was hospitalized for acute HF and underwent mechanical circulatory support followed by heart transplantation; the patient’s data were analyzed separately in the case study. 

A total of 22,551 audio recordings were collected from 32 patients (average 705 $\pm$ 59 per patient). After removing 311 outliers (1.38\%), 21,863 recordings (96.94\%) were retained for feature extraction, including 16,347 (511 $\pm$ 43) vowel and 5,516 (172 $\pm$ 15) speech recordings. Additionally, 1,816 weight and 3,632 blood pressure recordings were collected, with 6 weight (0.3\%) and 3 blood pressure (0.08\%) entries removed as outliers.

\begin{table}[htbp]
\centering
\begin{threeparttable}
\centering
\caption{Baseline Demographics and Clinical Characteristics (N = 32)}
\label{tab:clinical}
\small
\begin{tabularx}{\linewidth}{X p{4cm} p{4.5cm} p{2cm}}
\hline
\textbf{Variable} & \textbf{V1} & \textbf{V2} & \textbf{S} \\
\hline

Age (years) & $55.69 \pm 11.28$ & & \\

Sex & Male (90.62\%), Female (9.38\%) & & \\

Smoking & Stopped (78.12\%), Yes (12.5\%), Never (9.38\%) & & \\

Height (cm) & $173.78 \pm 9.23$ & & \\

Native German speaker & Yes (81.25\%), No (18.75\%) & & \\

Last LVEF (\%) & $27.89 \pm 9.07$ & & \\

Weight (kg) & $87.72 \pm 17.27$ & $86.72 \pm 16.82$ & $86.30 \pm 17.03$ \\

Systolic pressure (mm Hg) & $114.91 \pm 18.07$ & $116.00 \pm 17.93$ & $103.78 \pm 14.72$ \\

Diastolic pressure (mm Hg) & $69.52 \pm 10.74$ & $67.50 \pm 12.32$ & $65.82 \pm 10.20$ \\

Heart rate (bpm) & $68.28 \pm 13.56$ & $67.03 \pm 10.21$ & \\

NYHA class & II (84.38\%), III (15.62\%) & II (87.1\%), I (6.45\%), III (6.45\%) & \\

NT-proBNP (ng/L) & $493.0$--$2686.75$ & $426.5$--$1824.5$ & \\

Creatinine ($\mu$mol/L) & $120.53 \pm 40.22$ & $128.07 \pm 47.82$ & \\

HFaST & & & $15.93 \pm 7.83$ \\

KCCQ & & & $71.0 \pm 19.11$ \\

\hline
\end{tabularx}

\begin{tablenotes}
\footnotesize
\item Values are presented as $n$ (\%), mean $\pm$ SD, or interquartile range (IQR).
LVEF = left ventricular ejection fraction;
NT-proBNP = N-terminal pro B-type natriuretic peptide;
NYHA = New York Heart Association.
V1 and V2 refer to clinical visits before and after the longitudinal study, and S denotes measures collected during the study.
HFaST = Heart Failure Symptom Tracker.
\end{tablenotes}
\end{threeparttable}
\end{table}

\subsection{Repeated Measures Correlation}

\begin{figure}[htbp]
    \centering
    \includegraphics[width=\linewidth]{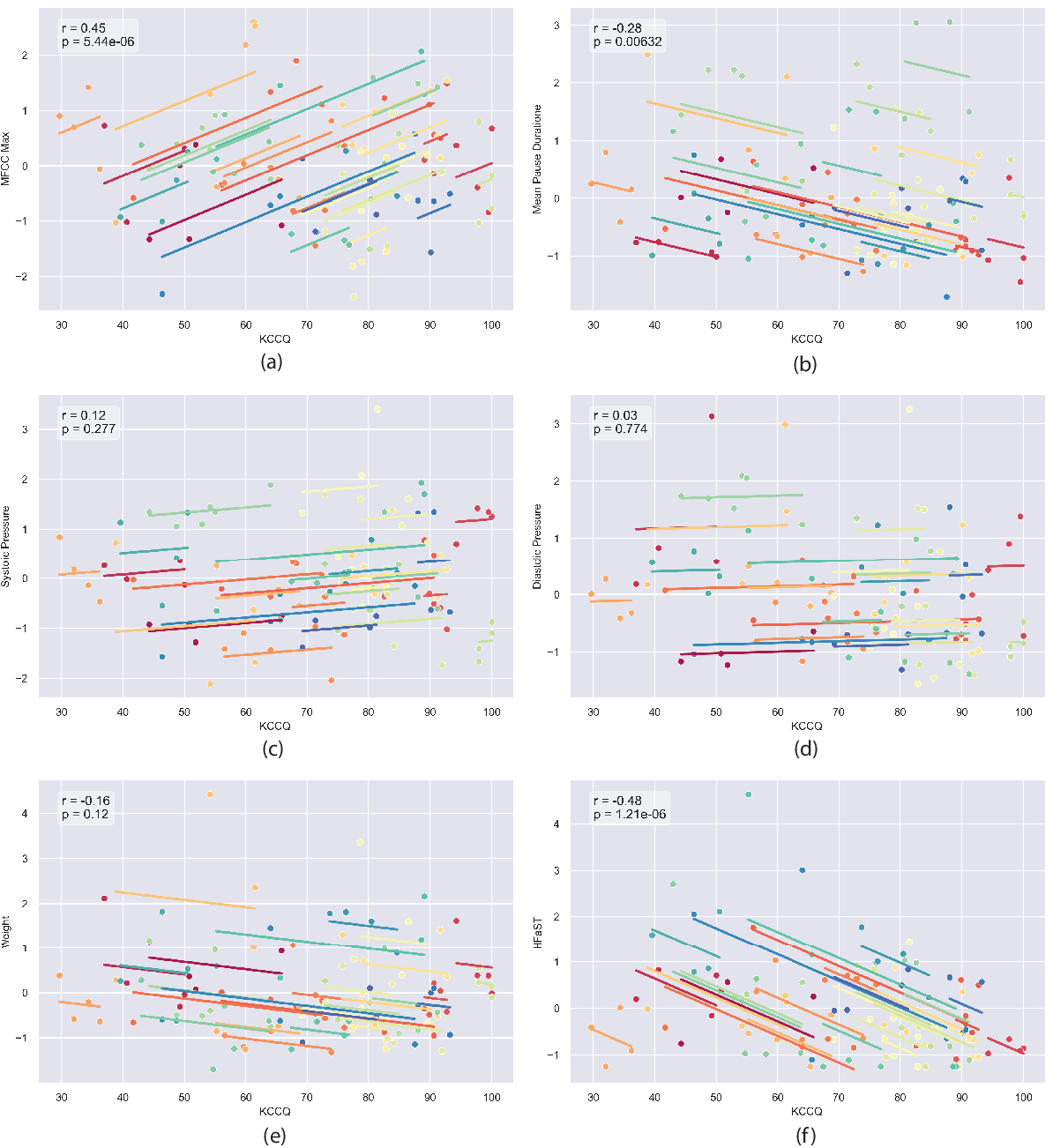}
    \caption{\textbf{Repeated measures correlation between health status and dynamic features}. Health status is represented by the KCCQ overall summary score. Shown are the standard deviations of selected features over a 7-day lookback window. Each color represents one patient. Panels include: (a) max MFCC of vowel /a/; (b) mean pause duration (command task); (c–d) blood pressure; (e) weight; (f) HFaST score. r indicates the correlation coefficient, and p the significance level. }
    \label{fig:fig1}
\end{figure}

Repeated-measures correlation showed that changes in vowel, speech, and HFaST scores were more strongly associated with the KCCQ summary score than SoC. Using a 7-day lookback window as an example, the standard deviation of vowel features demonstrated the strongest correlation with KCCQ ($|r| > 0.4$), speech features showed moderate correlation ($|r| \approx 0.2$), and weight and systolic pressure showed weaker correlations ($|r| = 0.16$ and $0.12$, respectively). The HFaST questionnaire showed the highest correlation with KCCQ ($|r| = 0.48$) (Figure~\ref{fig:fig1}).

\subsection{Classification}

\begin{figure}[htbp]
    \centering
    \includegraphics[width=\linewidth]{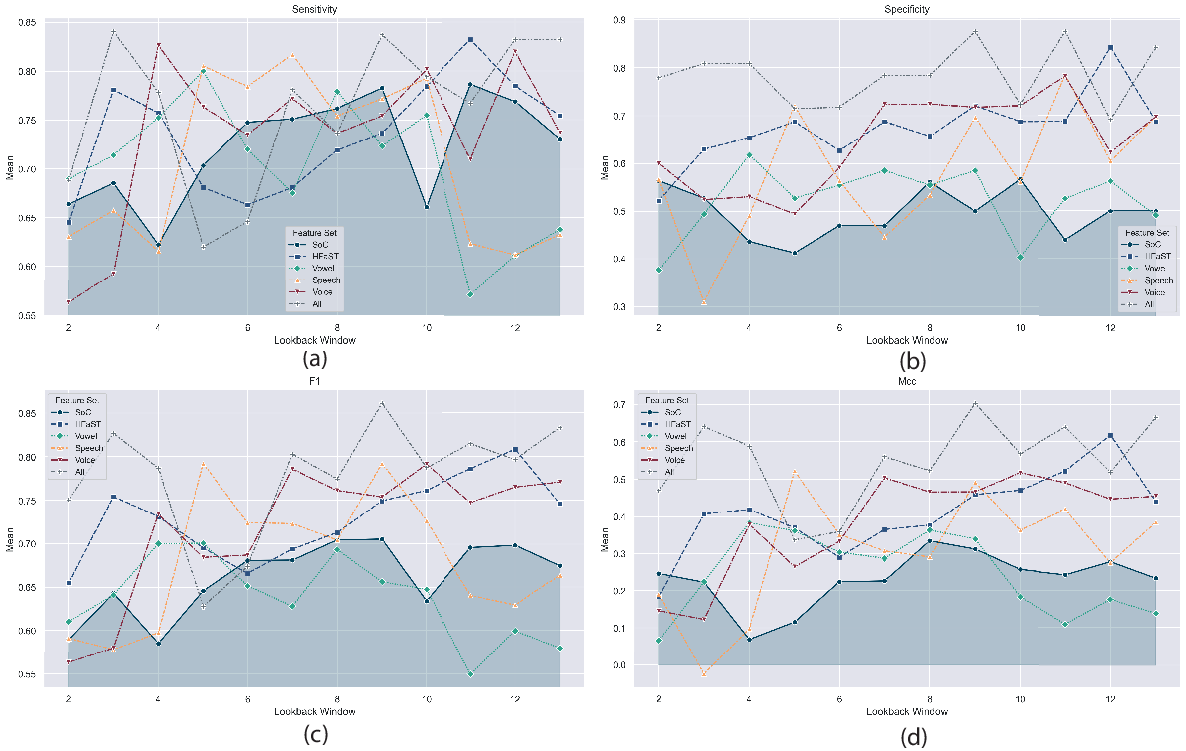}
    \caption{\textbf{Classification performance comparison across different feature sets.} (a) Sensitivity, (b) Specificity, (c) F1 score, and (d) MCC are reported. Each dot represents the mean performance of a feature set across multiple measurements for a given lookback window (K = 2–14 days). The shaded area shows the SoC performance, used as the baseline for comparison.}
    \label{fig:fig2}
\end{figure}

Figure~\ref{fig:fig2} shows Random Forest performance across lookback windows (K=2-14). The combined voice feature set (vowel + speech) consistently outperformed SoC, achieving higher MCC in 83.3\% of windows. Peak performances included sensitivity 0.826 (K = 4), specificity 0.782 (K = 11), F1 0.792 (K = 10), and MCC 0.518 (K = 10). By contrast, SoC performance peaked at sensitivity 0.783 and F1 0.705 (K = 9), specificity 0.567 (K = 10), and MCC 0.334 (K = 8).

\begin{figure}[htbp]
    \centering
    \includegraphics[width=\linewidth]{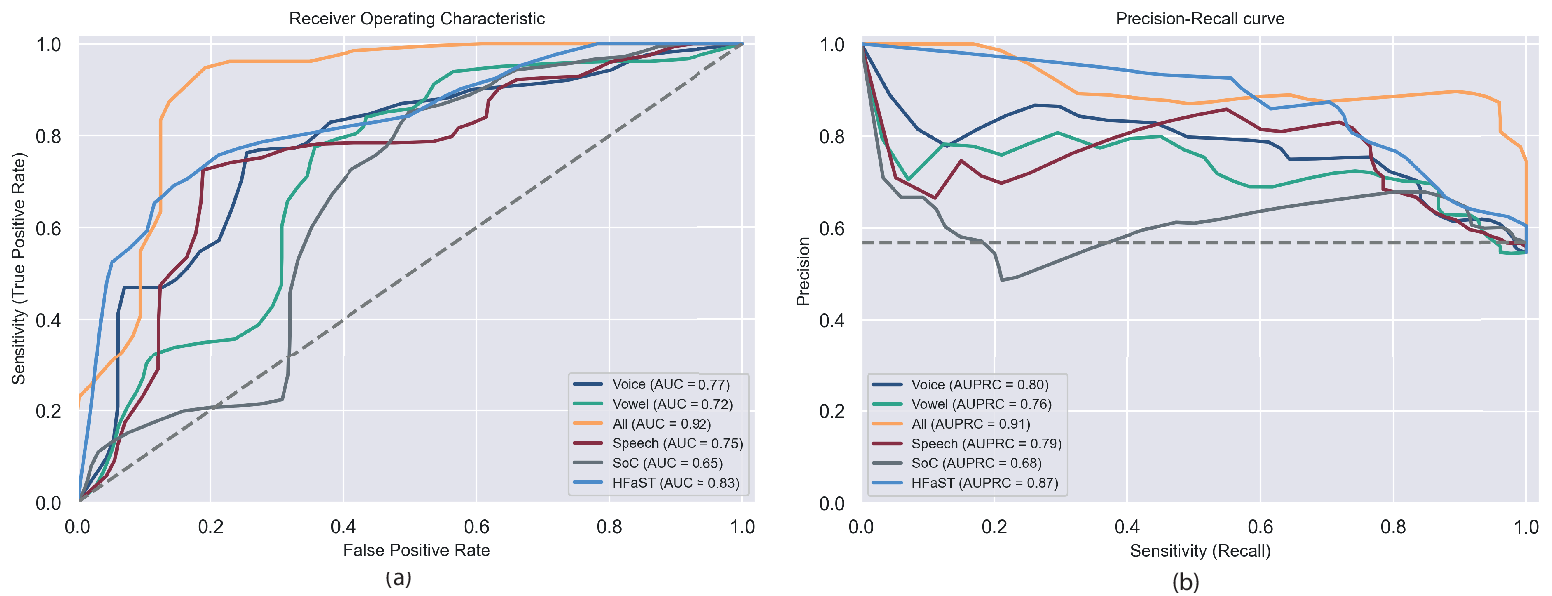}
    \caption{\textbf{ROC and Precision–Recall curves for different feature sets.} (a) AUC and (b) AUPRC are shown to indicate performance in distinguishing health deterioration. Each line represents a specific feature set using a lookback window of K = 9 days.}
    \label{fig:fig3}
\end{figure}

Among individual voice tasks, speech features generally outperformed vowels (peak sensitivity 0.817 vs. 0.800; peak specificity 0.785 vs. 0.618). The HFaST set showed steadily improving performance with longer windows, converging with Voice around K = 10.
The All feature set (SoC + HFaST + Voice) delivered the best results, with peak sensitivity 0.837, specificity 0.876, F1 0.861, and MCC 0.703 at K = 9, highlighting the robustness of multimodal integration. As shown in Figure~\ref{fig:fig3}, under the same window, Voice features showed better discrimination than SoC, with an AUC of 0.77 versus 0.65. Similarly, the AUPRC was higher for Voice (0.80) compared with SoC (0.68), indicating superior performance in identifying deterioration. Notably, both pure Vowel and Speech features also outperformed SoC, with AUCs of 0.72 and 0.75, and AUPRCs of 0.76 and 0.79, respectively. Multi-class classification performance for distinguishing intermediate health states is provided in the Supplementary Appendix.

\begin{figure}[htbp]
    \centering
    \includegraphics[width=\linewidth]{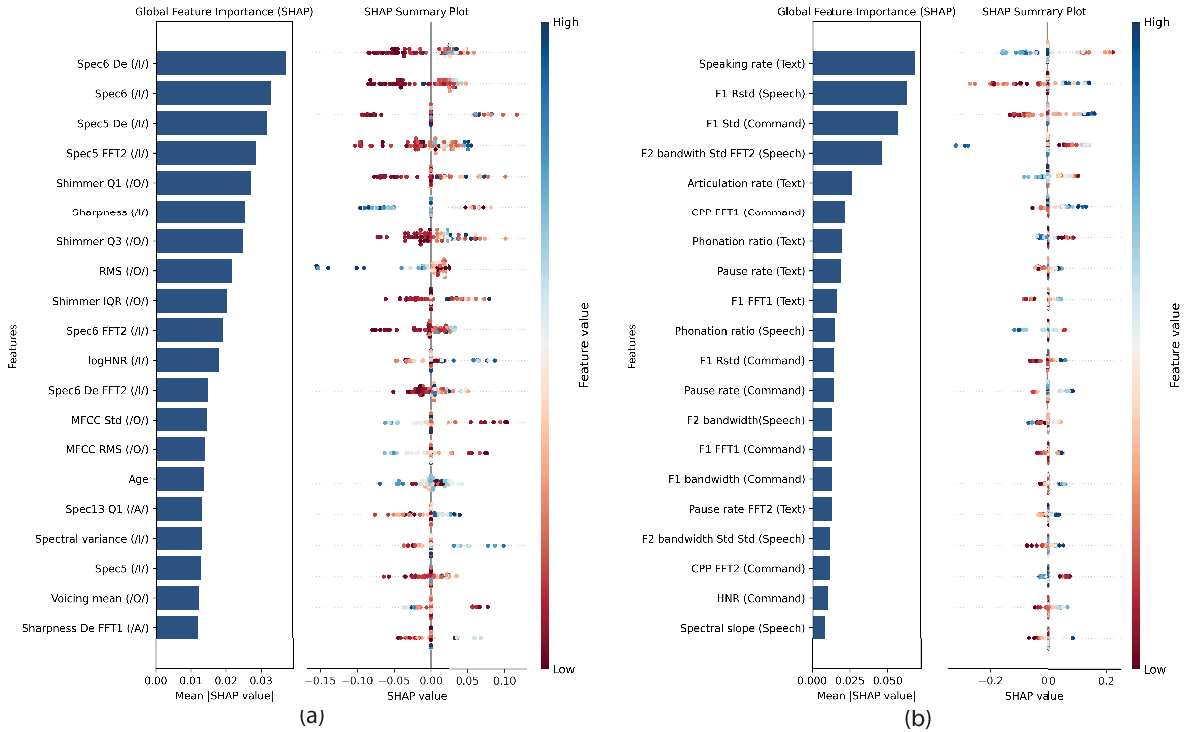}
    \caption{\textbf{SHAP global and summary plots for the voice feature set.} The top vowel (a) and speech (b) features are shown, with positive SHAP values indicating contributions toward worse health outcomes. Feature labels are simplified codes; see Supplementary Material for more descriptions.}
    \label{fig:fig4}
\end{figure}

Using a lookback window of K = 9, Figure~\ref{fig:fig4} highlights the top-ranked (coded) features from the Vowel and Speech sets that were most predictive for classifying health status. Full descriptions of each feature are provided in the Supplementary Material. For vowel features, poorer health was associated with slower or delayed energy peaks (mean over the lookback window) and greater shimmer variability and range, reflecting less stable vocal fold vibrations. In contrast, greater variability in spectral features (top Fourier components over the lookback window) was linked to better health, indicating more dynamic and flexible vocal patterns. Speech features revealed that higher speaking and articulation rates, along with greater phonation ratio, corresponded to better health, reflecting more fluent and efficient speech. In contrast, increased formant variability (frequency and bandwidth) indicated poorer health due to unstable articulation, whereas higher cepstral peak prominence signaled clearer and more stable voice quality.

\subsection{Statistical Analysis and Case Study}

When comparing the worst and best health states, we identified the most informative features reflecting deterioration, as shown in Table~\ref{tab:top_features_health_states}. Among speech features, phonation ratio, speaking rate, and articulation rate extracted from the reading text task were lower in worse health ($p < 0.001$, $r > 0.65$), while pause rate was higher ($p < 0.001$, $r < -0.75$). First formant variability from the command task was higher in worse health ($p < 0.001$, $r > 0.60$). For vowel features, shimmer of /o/ showed increased variability and the energy peak of /i/ occurred later in worse health ($p < 0.001$, $r > 0.50$).

\begin{table}[htbp]
\centering
\begin{threeparttable}
\centering
\caption{Statistical comparison of top features between best and worst health states.}
\label{tab:top_features_health_states}

\begin{tabular}{l c c c}
\hline
\textbf{Feature} & \textbf{Best states} & \textbf{Worst states} & \textbf{Effect size} \\
\hline

Phonation ratio (Text) & 0.914 (0.048) & 0.826 (0.158) & 0.846 \\
Speaking rate (Text) & 4.061 (0.231) & 3.347 (0.958) & 0.805 \\
F1 Std (Command) & 16.771 (6.895) & 29.279 (15.859) & 0.702 \\
Articulation rate (Text) & 4.446 (0.325) & 4.078 (0.644) & 0.662 \\
Shimmer Q3 (/O/) & 8.309e$-$07 (7.032e$-$07) & 2.000e$-$06 (2.771e$-$06) & 0.638 \\
F1 FFT1 (Command) & 70.583 (39.714) & 118.773 (63.202) & 0.626 \\
Spec6 (/I/) & 0.550 (0.192) & 0.848 (0.782) & 0.622 \\
Shimmer IQR (/O/) & 1.778e$-$06 (1.331e$-$06) & 4.081e$-$06 (5.516e$-$06) & 0.597 \\
Spec6 FFT2 (/I/) & 4.953 (1.915) & 7.342 (5.922) & 0.588 \\
Spec5 (/I/) & 0.543 (0.160) & 0.722 (0.704) & 0.580 \\
Spec6 De (/I/) & 0.608 (0.264) & 1.076 (0.901) & 0.576 \\
F1 Rstd (Speech) & 29.124 (11.612) & 43.677 (34.459) & 0.573 \\
Spec5 De (/I/) & 0.521 (0.206) & 0.784 (0.732) & 0.571 \\
Shimmer Q1 (/O/) & 9.411e$-$07 (7.030e$-$07) & 2.086e$-$06 (2.659e$-$06) & 0.568 \\
F1 bandwidth (Command) & 118.394 (86.329) & 188.749 (65.326) & 0.561 \\
Spec5 FFT2 (/I/) & 4.887 (1.608) & 6.383 (5.629) & 0.552 \\
Spec6 De FFT2 (/I/) & 5.468 (2.496) & 9.682 (7.316) & 0.543 \\
F1 Rstd (Command) & 16.177 (8.143) & 29.205 (25.539) & 0.525 \\
Spectral slope (Speech) & 5.661 (2.621) & 8.011 (3.909) & $-0.745$ \\
Pause rate FFT2 (Text) & 1.396 (0.435) & 1.908 (1.396) & $-0.760$ \\
Pause rate (Text) & 0.155 (0.052) & 0.246 (0.162) & $-0.788$ \\

\hline
\end{tabular}

\begin{tablenotes}
\footnotesize
\item Comparison of features between the worst and best health states. Values are reported as median (IQR). All listed comparisons were statistically significant ($p < 0.001$). Feature names are simplified codes; full descriptions are provided in Supplementary Material.
\end{tablenotes}
\end{threeparttable}
\end{table}

\begin{figure}[htbp]
    \centering
    \includegraphics[width=\linewidth]{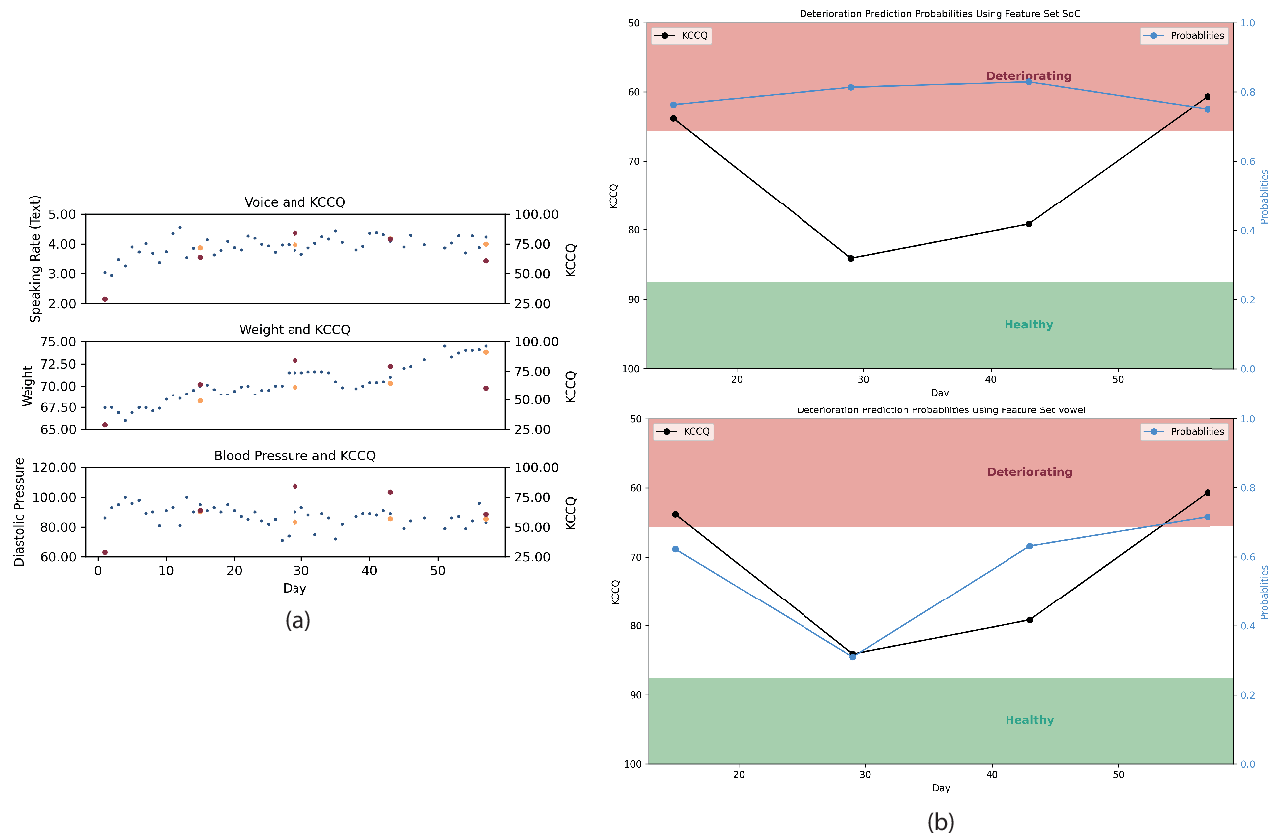}
    \caption{\textbf{Feature illustration and deterioration prediction for the case study patient.} (a) Key feature fluctuations (weight, diastolic pressure, text speaking rate) with mean values per lookback window K = 9 days (orange) and KCCQ scores (dark red). (b) Predicted deterioration probabilities (blue) versus KCCQ trajectory (black), with deteriorating (red) and healthy (green) regions; KCCQ axis reversed so upward indicates deteriorating.}
    \label{fig:fig5}
\end{figure}

In the case study, SoC features did not reflect deterioration trends, as predicted probabilities stayed high and nearly unchanged, as shown in Figure~\ref{fig:fig5}. In contrast, vowel features predicted the patient’s health status progression.

%%==================================%%
%% DISCUSSION
%%==================================%%

\section{Discussion}

\textbf{Key Findings}

Our study demonstrates the potential of vocal biomarkers for predicting health status in chronic HF. In 32 patients monitored longitudinally at home using mobile devices, voice features outperformed traditional SoC measures with higher sensitivity and specificity. Vowel features required shorter lookback windows (4–6 days), while speech features performed best over longer durations (5–10 days). Multimodal integration further improved robustness. Finally, a case study confirmed that voice features outperformed SoC in detecting health deterioration.

We identified key vowel and speech features that captured meaningful differences during health deterioration. Among vowel features, all three physiological categories were informative: vocal source (e.g., shimmer), vocal tract (e.g., spectral features, MFCCs), and prosody (e.g., energy). In particular, delayed energy shifts reflected vocal strain, fatigue, or breathlessness, which are linked to worse vocal function and articulation. On the other hand, low energy deviation indicates a strained, monotonous, or less dynamic voice, potentially reflecting fatigue or vocal strain. In the speech features, we noticed that a high speaking rate, articulation rate, phonation ratio, and cepstral peak prominence are indicators of more fluent and efficient speech production, which generally correlates with better health. On the other hand, greater variability in formant locations and bandwidths suggests unstable speech production, which can indicate vocal strain or fatigue, resulting in poorer health outcomes. Additionally, since the features are interdependent and interact with each other, decision-making based on voice should involve considering multiple features together, rather than relying on the cutoff values of individual features. Additionally, age emerged as an important predictor, as older individuals tended to show more subtle changes in their voice that could be related to overall health decline.

\textbf{Comparison with Previous Studies}

Previous studies have used weight and diastolic pressure for telemonitoring in acute HF admissions, achieving a best AUC of 0.82 $\pm$ 0.02~\cite{koulaouzidis2016telemonitoring}. In comparison, our best AUC of 0.77 using SoC feature set suggests that our weight and blood pressure can serve as a baseline when compared to voice features. In previous work using vocal biomarkers for HF, Schöbi et al. found that the pause ratio increased in acute patients with HF compared to patients with chronic HF~\cite{schobi2022evaluation}, which aligns with our finding that a higher phonation ratio indicates better health. Kiran et al. achieved an accuracy of 95.02\% in identifying health status in patients with HF using MFCC and glottal features~\cite{reddy2021automatic}, while Murton et al. reached up to 69\% accuracy in distinguishing admission and discharge in acute patients with HF using features such as cepstral peak prominence, pitch, and maximum phonation time~\cite{murton2023acoustic}. Our voice features achieved a peak sensitivity of 82.6\% and a peak specificity of 78.2\%, which is competitive with prior work, especially considering that our chronic patients, making health status detection more challenging than classifying health versus HF or admission versus discharge.

\textbf{Limitations}

We acknowledge several limitations in our study. First, the sample size was small, as we focused on patients with chronic HF without major comorbidities and required strict two-month daily participation, which limited recruitment. Second, the cohort lacked linguistic diversity, as all participants were German speaking. This may restrict the generalizability of our findings to other languages and cultural settings. Furthermore, there is a notable sex bias, with male patients being significantly more represented than female patients, which is a consequence of the difficult recruitment process. Therefore, validation in larger cohorts is needed. Regarding the home recordings, the absence of standardization in factors like recording distance, timing, and posture introduces variability, compounded by the patients' unfamiliarity with mobile devices. Besides, there are other potential confounders, such as emotional state or illness (e.g., a cold), that could affect voice recordings. Lastly, as patients became more familiar with the speaking task over the two-month period, their speaking rate tended to increase. In future studies, researchers can incorporate an additional text to address this learning effect. We also recommend including additional feature maximum phonation time.

\section{Conclusions}

This study shows that voice-based biomarkers, captured with mobile devices, can non-invasively predict health deterioration in patients with chronic HF more accurately than weight or blood pressure. The identified prognostic vocal biomarkers reflected subtle changes in health status, highlighting their potential for earlier detection of worsening HF and timely clinical intervention.

\bibliographystyle{unsrtnat}
\bibliography{references}  %%% Uncomment this line and comment out the ``thebibliography'' section below to use the external .bib file (using bibtex) .





% \end{thebibliography}

\end{document}